\documentstyle[aps,preprint,psfig,floats]{revtex}

\begin{document}

\draft
\tighten

%Define symbol for fractions:
\def\tfrac#1#2{{\textstyle{#1\over#2}}}

\title{Discreteness-induced resonances 
and AC voltage amplitudes\\
in long one-dimensional Josephson junction arrays}

\author{
A.\ E.\ Duwel$^{\mbox{\scriptsize a}}$,
Shinya Watanabe$^{\mbox{\scriptsize b}}$,
E.\ Tr\'{\i}as$^{\mbox{\scriptsize a}}$,\\
T.\ P.\ Orlando$^{\mbox{\scriptsize a}}$,
Herre S.\ J.\ van der Zant$^{\mbox{\scriptsize c}}$,
and Steven H.\ Strogatz$^{\mbox{\scriptsize d}}$
}

\author{\footnotesize
$^{\mbox{\scriptsize a}}$Dept.\ of Electrical Engineering \& Computer Science,
M.I.T., Cambridge, MA, 02143, USA\\
$^{\mbox{\scriptsize b}}$Ctr.\ for Chaos \& Turbulence Studies,
Niels Bohr Institute, Blegdamsvej 17, Copenhagen, 2100, Denmark\\
$^{\mbox{\scriptsize c}}$Dept.\ of Applied Physics, Delft University of
Technology, P.O.Box 5046, 2628 CJ Delft, The Netherlands\\
$^{\mbox{\scriptsize d}}$Dept.\ of Theoretical \& Applied Mechanics,
Cornell University, Kimball Hall, Ithaca, NY, 14853, USA
\vskip 3mm
to appear in the Journal of Applied Physics, autumn, 1997
}

\maketitle

% ********************************************************************

\abstract{%
New resonance steps are found in the experimental current-voltage
characteristics of long, discrete, one-dimensional Josephson junction 
arrays with open boundaries and
in an external magnetic field.
The junctions are underdamped, connected in parallel,
and DC biased.
Numerical simulations based on the discrete sine-Gordon model
are carried out,
and show that the solutions on the steps are periodic trains of
fluxons, phase-locked by a finite amplitude radiation.
Power spectra of the voltages consist of
a small number of harmonic peaks,
which may be exploited 
for possible oscillator applications.
The steps form a family that can be numbered by the harmonic content
of the radiation, the first member corresponding to the Eck step.
Discreteness of the arrays is shown to be essential for appearance
of the higher order steps.
We use a multi-mode extension of the harmonic balance analysis,
and estimate the resonance frequencies, the AC voltage amplitudes,
and the theoretical limit on the output power on the first two steps.
}


\section{Introduction}

Josephson junction systems have a natural application as millimeter-
and submillimeter-wave oscillators.  To facilitate their use, however,
ongoing research must address several issues.  To produce oscillators
with narrow linewidths and high power, we must study the harmonic 
content and oscillation amplitudes of Josephson sources.
Furthermore, the conditions under which many Josephson oscillators can
be phase-locked  to produce higher output power continue to
challenge researchers.

AC power has been measured from both continuous long Josephson junctions
and discrete arrays of short junctions. 
In underdamped Josephson systems, the AC oscillation amplitudes are expected
to be largest at certain resonant frequencies.  Much analytical
work has been done to predict the frequency of 
these resonances. The success of these
analyses can be assessed by studying the DC current-voltage ($I$--$V$)
characteristics.  Steps appear in the $I$--$V$ when increases of the current
bias over a certain range do not produce increases in the DC voltage.  Instead,
the input power drives large-amplitude AC oscillations. 
The frequency of these oscillations is related to the DC voltage
by the Josephson relation, $\omega =2 \pi V/ \Phi_o$, where $\Phi_o$ 
is the flux quantum, while the spatial wavenumber
is determined by the applied magnetic field, $k= 2 \pi f$, where
$f$ is the applied flux per unit cell normalized to $\Phi_o$.  The oscillations
can be approximated as the normal modes to the linearized system. 
This approach has been used successfully to predict resonance frequencies,
or steps in the $I$--$V$
characteristics of many different geometries of Josephson systems.  
The dependence of the step voltage on applied magnetic field can be included,
thus giving a dispersion relation for the associated oscillations.
In the discrete
case, the dispersion relation is nonlinear, and higher modes
waves are expected to produce distinct steps in the $I$--$V$ \cite{U:ring,Petr}.
One of the 
limitations of these linear analyses is that the predicted oscillations 
are necessarily
single-harmonic, since the normal modes of the linearized Josephson system
are simply Fourier modes.  
In addition, the absence of a driving force in the linearized
system precludes any calculation of the oscillation amplitudes. 

In order to predict oscillation amplitudes  and the distribution of
power among excited modes, the nonlinearity in the system must be more
carefully treated.
Perturbation techniques \cite{s:whirl,yoshida}  clarify the role of the
nonlinearity in driving the resonance and have been used to predict step
voltages in Josephson systems.
Combined with an appropriate ansatz, they also provide expressions for 
the oscillation amplitude 
\cite{jap96,kulik,enpuku}.  However, analytic expressions for the AC 
amplitudes so far include only the first harmonic and are implicit 
\cite{jap96}.  

In this paper we use experiments, simulations, and non-linear
analysis to investigate
the properties of discrete, planar arrays of Josephson junctions 
connected in parallel. 
Both experiments and simulations yield several resonant steps in the $I$--$V$.
Simulations indicate that on these steps, a traveling wave pattern dominates
the oscillations of junctions in the array. 
In contrast to the linear picture, 
these resonances correspond to  excitations
of more than one  mode, though the harmonic content is still limited.  
Similar states have already been reported in discrete rings \cite{kink}, 
indicating that boundary conditions play a minor role.  
In inductively coupled arrays with open ends, 
the second harmonic resonance was also observed experimentally \cite{asc96}. 
We use a two-mode extension of the harmonic balance method
to predict the resonant frequencies as well as the
mode amplitudes at resonance.
Our formulas are analytic and include the dependence on magnetic field.
With a matched load condition, a theoretical upper
limit on the available output power of the underdamped Josephson oscillator
is calculated.

% ********************************************************************

\section{Measurements of Arrays}

\begin{figure}[tbp]
\centerline{\psfig{file=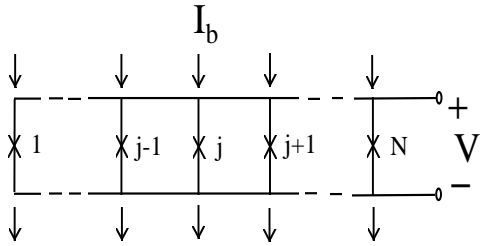,height=1.5in,width=2.75in}}
\caption{Schematic of an open-ended parallel array. A uniform current
$I_b$ is applied at each of the upper nodes and extracted at the lower
nodes.} 
\label{fig:array}
\end{figure}

We have measured single-row arrays of $N=54$ junctions 
connected in parallel.
Figure~\ref{fig:array} shows a schematic of our device.
A bias current $I_b$ is applied at each upper node of the array
and extracted from each bottom node as shown. We use resistors to
distribute the bias current as evenly as possible.
The voltage across the array is measured at an edge.
The array is placed above a superconducting ground plane, 
and a separate control line (not shown) is
used to apply a magnetic field. 
We will discuss the applied field in terms of  the {\em frustration}, $f$
(the flux applied to a single loop of the array,
normalized to the flux quantum).
Because the system is discrete,
we expect its properties to be 
periodic in $f$ with 
period $f=1$ \cite{Herre:Eck}. 
In this experiment, the applied flux is
proportional to the control current.

Samples were fabricated using a Nb trilayer process \cite{hypres}.
The junctions are $3 \times 3 \, \mu {\rm m}^2$ 
with a critical current density
of $j_c(T=0)= 1270 \, {\rm A/cm}^2$.  Device parameters have been
determined using the diagnostic procedures described by van der Zant
{\it et al.} \cite{diag}.  
For our samples at $7.2 \,$K, 
the normal-state resisitance $R_n=16.6 \, \Omega$,
the self-inductance of a loop $L_s= 6.4 \,$pH, the nearest-neighbor
inductive coupling $M_h=0.11L_s$, the junction capacitance
$C=340 \,$fF, and the Josephson inductance 
$L_J= \Phi_o / (2 \pi I_c) = 5.9 \,$pH. 
We  use the normal-state resistance to calculate the Stewart-McCumber
parameter, $\beta_c=16$. 
The sub-gap resistance is not as well defined at
these high temperatures but approaches the value of $R_n$. 
For the discretness
parameter, we calculate $\Lambda_J^2 = L_J / L_s =0.92$. 

\begin{figure}[tbp]
\centerline{\psfig{file=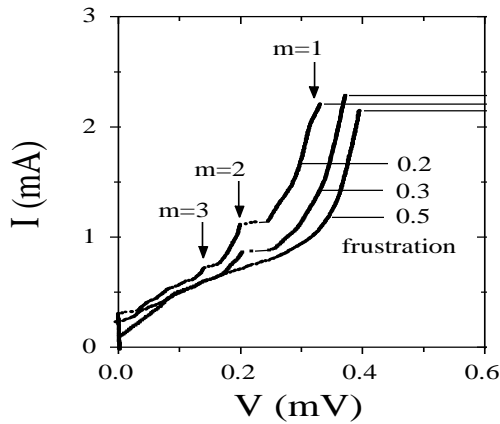,height=2.5in,width=2.75in}}
\caption{Current vs.\ voltage of a 54-junction array on a ground plane
at three values of $f=0.2,0.3,0.5$.
The temperature is $7.2 \,$K so that $I_cR_n=0.93 \,$mV.
The three steps indicated by the arrows are labeled by $m$ values
corresponding the number of dominant harmonics in the mode.}
\label{fig:ivs}
\end{figure}

Figure~\ref{fig:ivs} shows the voltage across an array
when the current is uniformly
injected. Three $I$--$V$ curves are presented, 
for $f=0.2$, $0.3$, and $0.5$.  
The most prominent feature is the sharp Eck step, 
which is the steepest part of the $I$--$V$, just before the 
switch to the gap voltage (not shown).
We label this step ``$m=1$'' for the reason
given in the next section.
In the regime where $\beta_c$ is small (overdamped)
or $\Lambda_J$ is large (less discrete),
this is the only step observed.
For our underdamped and discrete samples, however,
additional steps appear below the Eck voltage.
We index these  steps with $m=2, 3$, etc.
In this paper we are mostly concerned with the Eck step $m=1$ and,
among the new steps, the clearest one with $m=2$.
The study of the second step sheds light on the finer steps with $m>2$.

\begin{figure}[tbp]
\centerline{\psfig{file=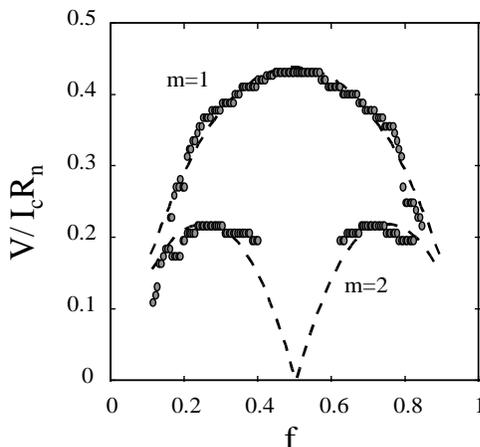,height=2.5in,width=2.75in}}
\caption{Step voltages of the 54-junction array vs.\ the frustration, $f$,
for the modes $m=1$ and $m=2$.  The dashed curves are plots of Equation
(13). }
\label{fig:vstep210}
\end{figure}

The voltage locations of the peaks experimentally vary with $f$,
as seen in Fig.~\ref{fig:ivs}.
More systematically, we show the dependence of the first two steps
in Fig.~\ref{fig:vstep210}.
The Eck peak voltage is found to be periodic in $f$
with period $f=1$
and to be approximately symmetric with respect to $f=0.5$.
This is consistent with previous observations \cite{Herre:Eck}.
At $f=0.5$, the Eck step reaches its highest voltage value.
For a smaller $f$, there is a threshold frustration, 
below which the Eck step does not appear.
This cut-off $f_{c1}$, 
known as the lower critical field or frustration,
is the minimum applied flux density for vortices 
to enter an array \cite{Herre:Eck}.
The value is quite large for our system, 
$f_{c1} \approx 2 / (\pi^2 \Lambda_J) = 0.2$.
The voltage location of the second step shows roughly the same
$f$-periodicity and symmetry as the Eck step.
This second step, however, achieves the maximum voltage near
$f=0.25$, and it
disappears near $f=0.5$ and for approximately $f<f_{c1}$.

The Eck ($m=1$) steps are ubiquitous in one-dimensional parallel arrays
as well as in continuous long junctions.
In contrast, the other steps ($m>1$) 
do not appear in long continuous junctions, but do appear
in discrete arrays when $\Lambda_J$ is small.
In our arrays, we find $\Lambda_J$ must be less than unity
for the $m=2$ step to appear.
Similar steps have been observed also in highly discrete
circular arrays \cite{kink} 
and open-ended arrays consisting of two rows 
that are inductively coupled \cite{jap96}.

In open-ended arrays with a smaller $N$,
Fiske steps \cite{Herre:Fiske} may be observed 
in a similar part of the $I$--$V$ below the Eck voltage.
We emphasize, however, that they are qualitatively different.
Fiske resonances can be described as standing waves (cavity modes) 
resulting from boundary reflections.
The wavelength of the modes are restricted by the boundary geometry,
and consequently, the resonance voltage locations 
do not depend strongly on $f$.
At a certain value of $f$, only even or only odd modes are excited.
As $N$ becomes large, for a given value of damping, 
these Fiske resonances disappear due to damping of the edge reflections. 
None of these features apply for the Eck step as well as
the $m>1$ steps,
which are tunable in $f$.
Thus, the new steps are   
expected to belong to the same family as the Eck step.

% ********************************************************************

\section{Simulations}

The governing equations which model our arrays are derived 
by applying Kirchhoff's
current laws and using the RSJ model for the current through a single
junction. We normalize the current to $I_c$, the voltage to $I_cR_n$,
and time to $\sqrt{L_JC}$ (inverse plasma frequency).
The equations are given in terms of the gauge-invariant
phase differences across the junctions,  $\phi_j$, where
$j=1,\dots,N$ indexes the junction's position.
For simplicity we neglect all the cell inductances 
except the self-inductance,
which results in the damped, driven, discrete sine-Gordon model:
\begin{equation}
  \ddot{\phi}_j + \Gamma \dot \phi_j + \sin \phi_j =
  I_b/I_c + \Lambda^2_J (\phi_{j+1} -2 \phi_{j} + \phi_{j-1} ) 
\label{phigov}
\end{equation}
for $j=1,\dots,N$ and $\Gamma = \beta_c^{-1/2}$.
Our numerical code can include
longer range inductances as necessary,
up to the full inductance matrix \cite{ETthesis}.
However, our analysis in the next section uses mainly (\ref{phigov}).
Simulating the same equations allows us to make a direct comparison 
with our analysis.
The simpler system not only illuminates the essential mechanism 
responsible for the observed steps
but also reproduces the measurements reasonably well,
as we will show in this section.

\begin{figure}[tbp]
\centerline{
\psfig{file=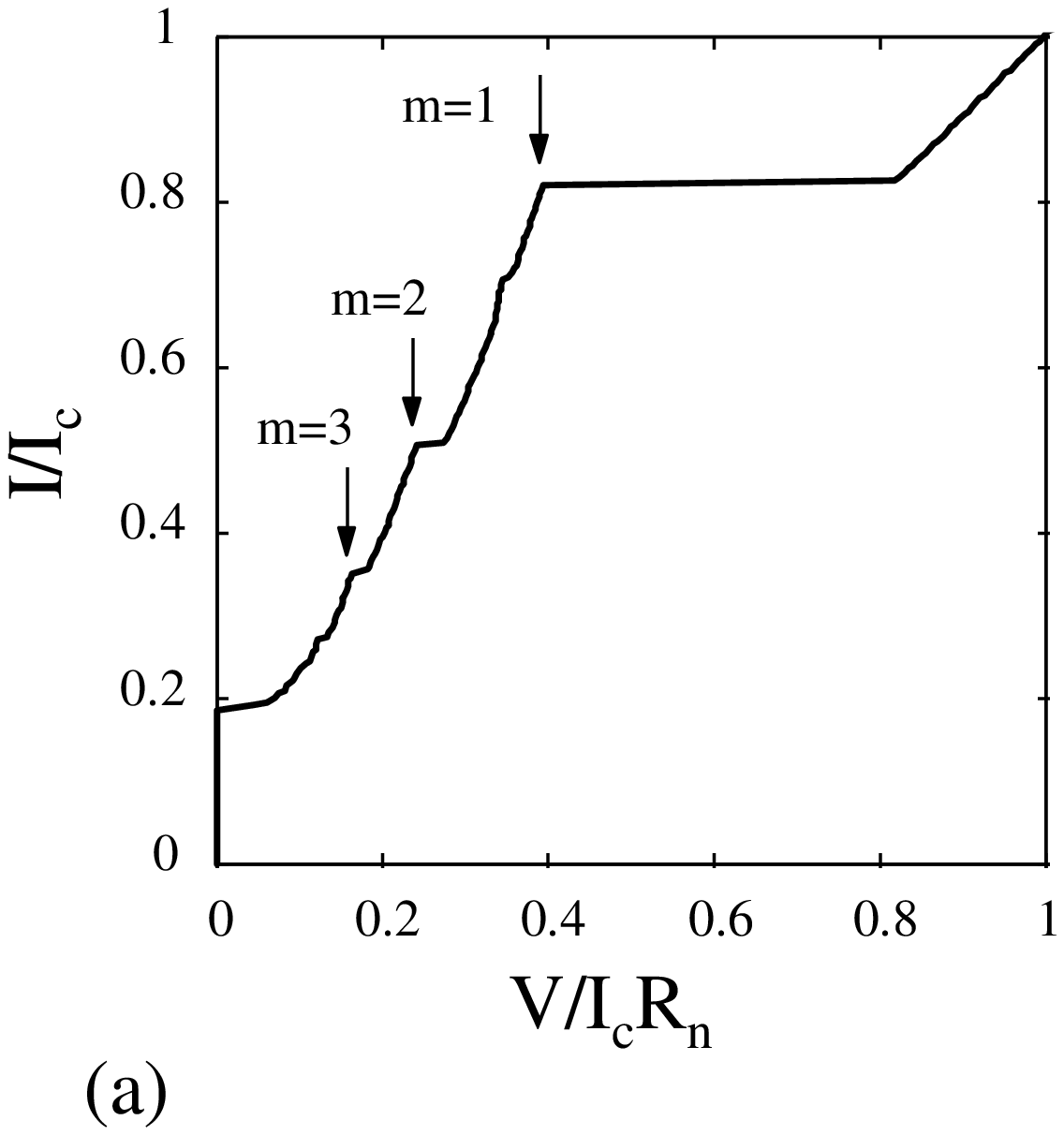,height=2.5in,width=2.5in}
\psfig{file=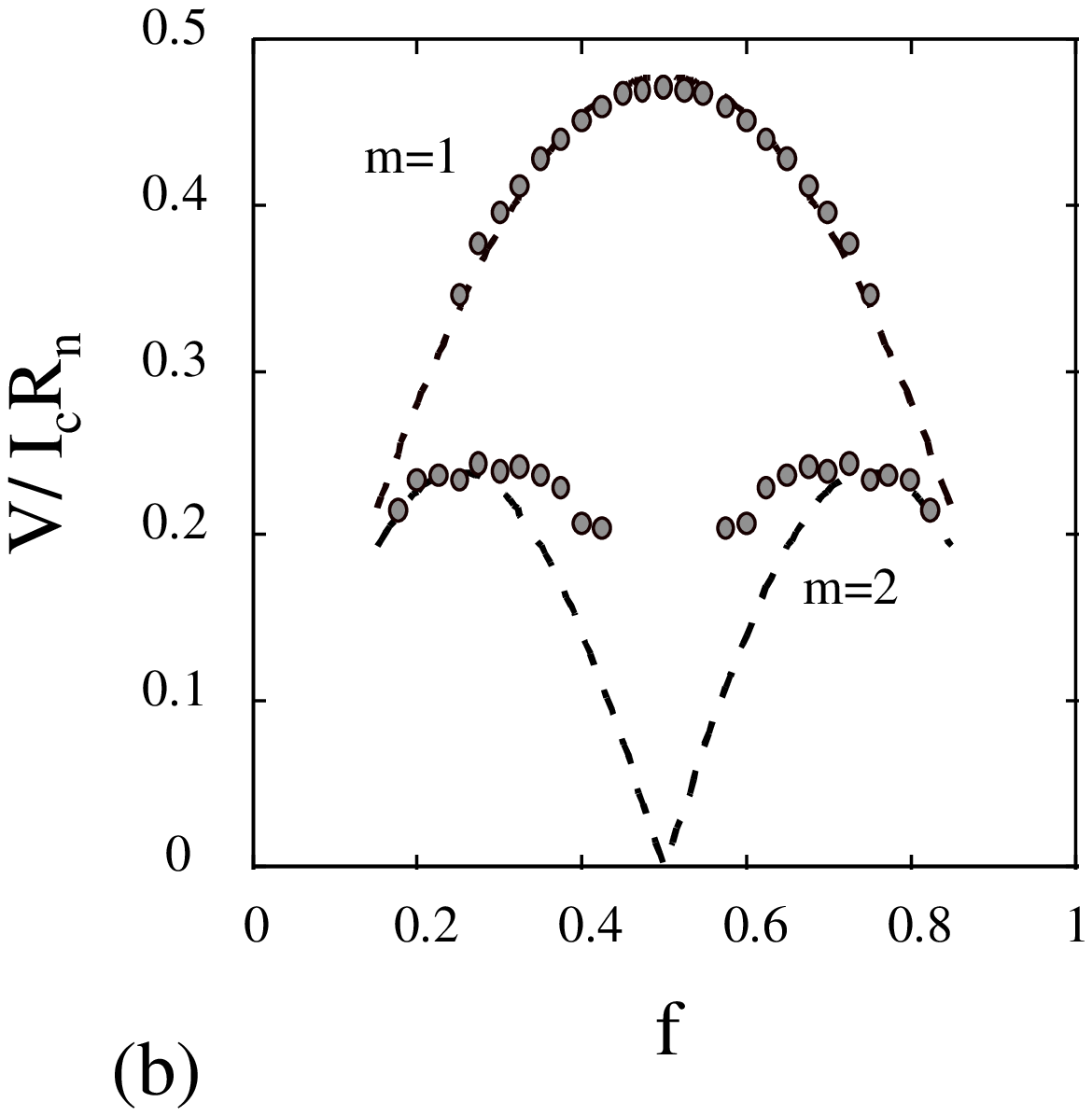,height=2.5in,width=2.5in}
}
\caption{(a) Simulated current-voltage characteristic for a
54-junction array at $f=0.3$.  The three most prominent steps are marked.
(b) Voltages corresponding to the top of the steps show the tunability of 
the first two steps with magnetic field.
The parameters are the experimental values:
$\beta_c=16$ ($\Gamma=0.25$) and $\Lambda_J^2=0.92$.}
\label{fig:simIV}
\end{figure}

With only self-inductance, the boundary conditions are simply
\begin{equation}
  \phi_0 (t) = \phi_1 (t) - 2 \pi f \qquad {\rm and} \qquad
  \phi_{N+1} (t) = \phi_N (t) + 2 \pi f
\label{boundarycondition}
\end{equation}
for all $t$, where artificial junctions
$\phi_0$ and $\phi_{N+1}$ are introduced at the endpoints
so that (\ref{phigov}) is valid at $j=1$ and $N$ as well \cite{s:longpaper}.
The fourth-order Runge-Kutta scheme
with a time-step $\Delta t = 1$ 
is used for integrating the system.
The instantaneous voltage at junction $j$ is 
simply proportional to the rate of the change of $\phi_j$,
and is given in our normalization by
\begin{equation}
  V/I_c R_n = \Gamma d\phi_j/dt .
\label{voltagefrequencyconversion}
\end{equation}

With (\ref{phigov}--\ref{voltagefrequencyconversion}),
current-voltage characteristics are numerically obtained
at different $f$ values 
using the parameters $\Lambda_J$ and $\Gamma$ from our experiments.  
Figure~\ref{fig:simIV}(a) shows the results for $f=0.3$.  
The curve reproduces the measured one in Fig.~\ref{fig:ivs}
well, and at least three steps (indexed by $m$) are clear.
(There is also a small step just below the $m=1$ step
which we think arises from a different mechanism.)
In a manner similar to Fig.~\ref{fig:vstep210},
the $f$-dependence of the step voltage locations
is shown in Fig.~\ref{fig:simIV}(b).
There are slight differences, but
the main features and voltage values are represented well.

\begin{figure}[tbp]
\centerline{\psfig{file=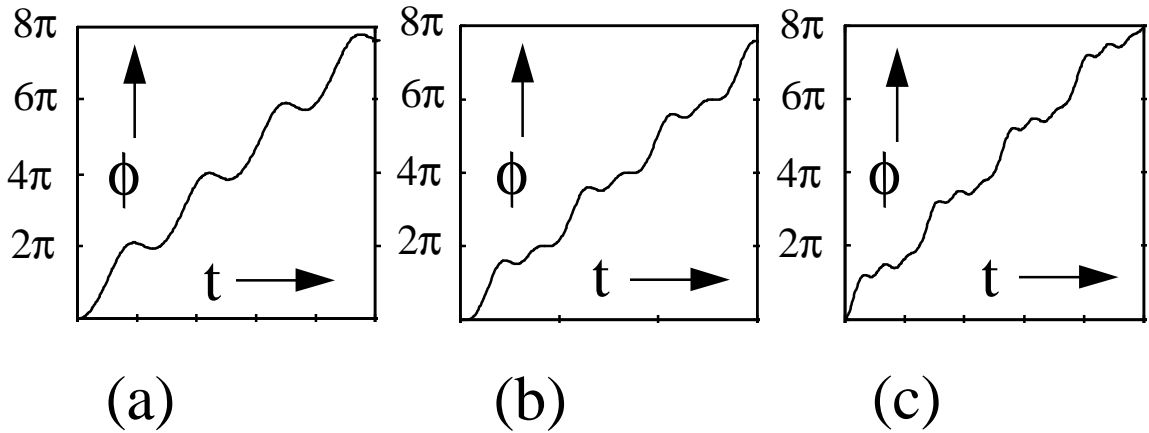,height=2.75in}}
\caption{Time evolution of the phase of a junction $j=27$
in the center of an array, biased on the
(a) $m=1$, (b) $m=2$, and (3) $m=3$ steps, respectively.
Note the increased frequency content as the mode number increases.}
\label{phase}
\end{figure}

Simulations allow us to study the solutions in detail.
We are especially interested in the system dynamics 
when biased on top of a step.
In Fig.~\ref{phase}(a--c)
the phase of the junction $j=27$ 
(located in the middle of the array)
on the $m=1,2,3$ step, respectively, is shown
as a function of time.
The junction appears to be in a periodic motion.
Its phase increases rapidly when a vortex (kink) passes by,
and it oscillates for the period between passing vortices.
In a mechanical analog of a junction as an underdamped pendulum \cite{steve},
a sudden overturn of the pendulum is followed by an overshoot,
and the pendulum ``rings'' several times until the
next vortex passes by.

\begin{figure}[tbp]
\centerline{\psfig{file=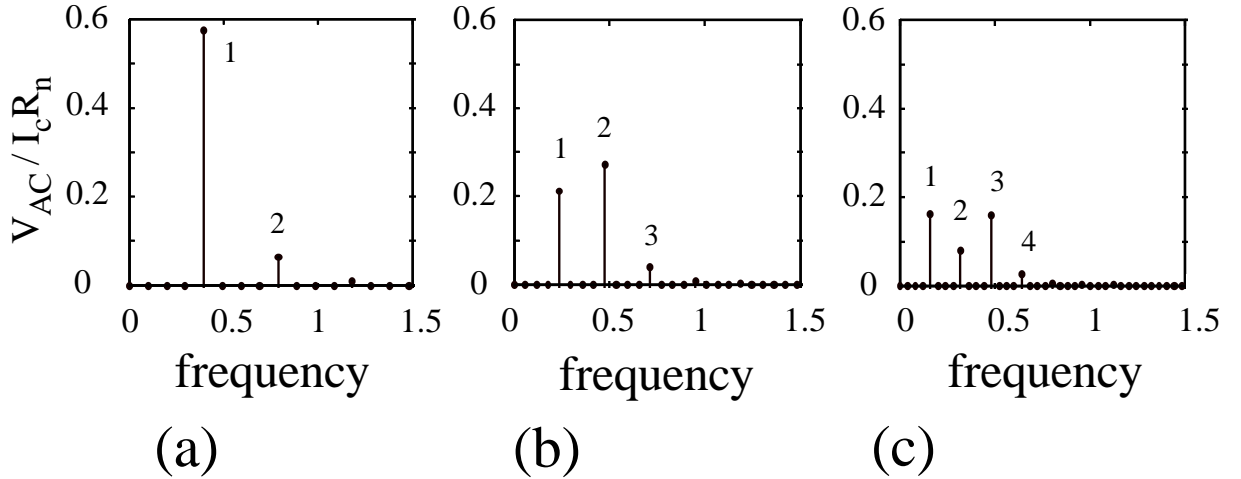,height=2.75in}}
\caption{Fourier spectra of the voltage 
corresponding to Fig.~\protect{\ref{phase}}(a--c), with
(a) step m=1, (b) step m=2, and (c) step m=3.
Voltages are proportional to the time derivative of the phase
in Fig.~\protect{\ref{phase}}. The average slope of the phase plots
corresponds to the DC voltage and is determined by the fundamental
harmonic.  The  DC voltage was subtracted  before
computing the Fourier spectra.  Note that only $m$ harmonics are 
dominant for the $m$th step.}
\label{spectra}
\end{figure}

It appears in Fig.~\ref{phase} that
the solution on the step $m$ corresponds to $m$ such ringings.
This can be quantified by studying the harmonic content 
of the voltages (\ref{voltagefrequencyconversion}).
Fourier spectra of the voltages are
shown in Fig.~\ref{spectra}(a--c), respectively.
On the step $m$, the first $m$ harmonics are dominant,
and the higher harmonics have rapidly decaying magnitudes.
Despite the presence of other harmonics, the steps are indexed by the 
ringing frequency, $m$.

Similar plots for the other junctions inside the array
appear identical to $\phi_{27}$,
except for a certain shift in the time axis.
This suggests that
the solutions are well approximated by traveling waves.
Near both ends of the array, 
reflections from the ends change this picture.
The boundary effects, however, decay
within 4--5 junctions from the end, 
and appear to play only a minor role
in our long arrays.

Such traveling wave solutions, consisting of a vortex
and ringing, have been found in circular arrays
with the periodic boundary conditions
\cite{U:ring,s:longpaper,StrunzElmer}.
In these systems, a vortex is trapped in a ring,
and as it circulates, it creates an oscillatory wake,
which phase-locks back to the vortex. 
It is not surprising that a nearly identical situation 
may arise in a linear array with open boundaries,
if the array is sufficiently long.
Instead of a single circulating vortex,
a vortex lattice propagates through the array.
A junction is swept periodically by a vortex in both cases.
There is, however, one major difference between the two geometries.
The magnetic field in the arrayin the array
(which controls the vortex spacing)
can be continuously tuned in the linear geometry
while it is restricted to multiples of $\Phi_0$
in a circular array due to the flux quantization.

%%%%%%%%%%%%%%%%%%%%%%%%%%%%%%%%%%%%%%%%%%%%%%%%%%%%%%%%%%%%%%%%%%%%%%%

\section{Analysis}

In this section we seek a more analytical description of the system,
and estimate DC voltages at each step and the amplitudes 
of the AC voltage components.
The simulations in the previous section suggest that
finding traveling wave solutions to the governing equation (\ref{phigov})
is the key to the estimates,
but this task is not simple in practice.

The equation (\ref{phigov}) may be viewed 
as a variation, in two respects,
from the {\em integrable} sine-Gordon partial differential equation,
\begin{equation}
  \phi_{tt} + \sin \phi = \phi_{xx}
\label{integrableSG}
\end{equation}
which possesses traveling kink solutions as exact solutions.
Firstly, (\ref{phigov}) is no longer a conservative equation,
as it includes the external drive (bias current $I_b$) 
and the loss ($\Gamma \dot{\phi_j}$) terms.
The added terms also break the integrability and 
exact solutions are no longer known.
A perturbation approach \cite{Parmentier93}
in the {\em conservative limit} ($I_b, \Gamma \rightarrow 0$)
has been developed to approximate a kink solution.

Secondly, the spatial derivative in (\ref{integrableSG})
is discretized in (\ref{phigov}).
In general, nonlinear wave equations discretized on lattices 
may exhibit qualitatively different solutions 
from their continuous counterparts,
and study of such discreteness-induced effects 
is an active current topic 
in its own right \cite{Duncan93}.
For  sine-Gordon systems, in particular,
(\ref{phigov}) was studied by \cite{Peyrard84,Willis,Duncan93}
without a loss term or a drive,
and it was found that propagation of a kink introduces 
a background radiation which greatly influences the speed of the kink.
As far as we are aware, however,
there has been no attempt to estimate the amplitudes
of the induced radiation, especially in the driven case.

Strunz and Elmer \cite{StrunzElmer} recently transformed
(\ref{phigov}) into a system of coupled modal equations and pointed out
that the superharmonic resonance leads to creation of
the radiated waves.
They expanded a traveling waveform into Fourier modes,
which may not be the best expansion basis but is a convenient one.
Since our Fourier spectrum in Fig.~\ref{spectra} shows
only a small number of peaks, 
we may truncate most modes and still obtain reasonable predictions.
With this goal in mind,
we review the analysis of \cite{StrunzElmer} in Sec.~4A.
The coupling terms among modes are truncated and analyzed in Sec.~4B,
and we estimate the AC voltage amplitudes on the steps.
The available power from the array is then evaluated in Sec.~4C.

\subsection{Resonance mechanism and voltages}

We look for traveling wave solutions of 
the discrete sine-Gordon equation (\ref{phigov}) of the form
\begin{equation}
  \phi_j(t) = \phi(x) = x + \psi(x)
\label{travelingwavesolution}
\end{equation}
where
\begin{equation}
  x = \omega t + 2 \pi f j
\label{eq:chi}
\end{equation}
is the moving coordinate with the wave, 
and $\psi(x+2\pi)=\psi(x)$.
The fundamental frequency $\omega$ is proportional to the DC voltage
through the Josephson voltage-phase relation 
(\ref{voltagefrequencyconversion})
while the spatial wavenumber is imposed by the external field.
If the modulation were absent ($\psi \equiv 0$), then
the boundary conditions (\ref{boundarycondition}) would be satisfied
exactly.
Since $\psi$ is not vanishing,
they are satisfied only on average,
and there should be a correction to (\ref{travelingwavesolution}).
We neglect this boundary effect
and will show that the simplification still leads to 
good estimates of the measurements and simulations.
(In a circular array (\ref{boundarycondition})
is replaced by the periodic boundary conditions:
$\phi_{j+N}(t) = \phi_j(t) + 2 \pi M$
where $M$ is an integer \cite{s:longpaper}.
There can be exact solutions of the form
(\ref{travelingwavesolution},\ref{eq:chi}) with $f=M/N$.)

The periodic function $\psi$ can be expanded into Fourier modes
\begin{equation}
  \psi(x) =  \sum_{m=-\infty}^{\infty} A_m e^{imx} 
\label{fourierexpansion}
\end{equation}  
with $A^*_{-m} = A_m$.
We set $A_0=0$, without loss of generality,
by shifting the origin of time.
The phase $\phi$ is an increasing function of $x$,
but the nonlinear term $\sin \phi$
is $2 \pi$-periodic and can be expanded as
\begin{equation}
  \sin \phi(x) = \sum_{m=-\infty}^{\infty} F_m e^{imx} 
\label{nonlinearityexpansion}
\end{equation}  
with $F^*_{-m} = F_m$.
The coefficients $F_m$ can be computed in terms of 
$A_{\pm 1}, A_{\pm 2}, \dots$
from the usual Fourier-Bessel expansions,
and therefore provide coupling among the modes.
By substituting (\ref{travelingwavesolution},\ref{fourierexpansion},\ref{nonlinearityexpansion})
into (\ref{phigov}),
we obtain a coupled system of modal equations.
\begin{equation}
  (\delta_m + i m \Gamma \omega) A_m + F_m = 0
\label{modaleqns}
\end{equation}
where
\begin{equation}
  \delta_m = \omega_m^2 - (m \omega)^2 ,
\label{delta}
\end{equation}
and
\begin{equation}
  \omega_m = 2 \Lambda_J \sin (m \pi f) .
\label{wm}
\end{equation}
The index $m=1,2,\dots$ for (\ref{modaleqns}--\ref{wm}).
In addition, the balance of the DC terms in (\ref{phigov}) results in
\begin{equation}
  I_b/I_c = \Gamma \omega + F_0 .
\label{dcmodaleqn}
\end{equation}
The first term on the right hand side is proportional
to $\omega$, and hence to the DC voltage.
This term describes the ohmic line in the $I$--$V$ plane.
The second term is also a function of $\omega$ 
through the $A_m$'s, and
describes the deviation of the $I$--$V$ curve from the ohmic
the line.

The superharmonic resonance \cite{Neyfeh} may occur
in the algebraic system (\ref{modaleqns},\ref{dcmodaleqn})
and lead to the creation of resonanct steps in the $I$--$V$ curve.
For small $\Gamma$, the magnitude of $A_m$ would generically
become large near $\delta_m=0$, that is, near
\begin{equation}
  \omega = \omega_m / m , \qquad m = 1,2,\dots .
\label{resonancefrequency1}
\end{equation}
Then $F_0$ also becomes large,
and the $I$--$V$ deviates substantially from the
ohmic line near these frequencies (or corresponding voltages).
This argument neglects the dependence of the coupling terms $F_m$
on the complex amplitudes $A_m$, but explains our measurements
quite well.
We plot (\ref{resonancefrequency1}) for $m=1,2$ as dashed curves 
in Figs.~\ref{fig:vstep210} and  \ref{fig:simIV}(b).
The agreement is good without a fitting parameter.

The right hand side of (\ref{resonancefrequency1}) 
is the phase velocity of the $m$th mode on the discrete lattice.
The left hand side can be viewed as the vortex velocity. 
Then, the relation can be viewed as the phase-locking condition
of the vortex velocity and the phase velocity of the 
$m$th mode \cite{U:ring,s:longpaper}.
In the absence of the coupling terms $F_m$,
a resonance occurs when the vortex lattice moves
at the same velocity as one of the modes.
We note that these resonance frequencies $\omega$ are distinct
for different $m$ because of the sinusoidal dependence of
the dispersion relation (\ref{wm}) on $m$.
If the second order difference in (\ref{phigov}) were
replaced by the second spatial derivative as in (\ref{integrableSG}),
the dispersion relation would depend linearly on $m$,
and the phase velocities of all the modes would be identical.
Consequently, there would be only one vortex velocity that can
excite modes, and we would observe only one resonance voltage.
This explains why only the Eck step 
(among this type of step)
is observed in a long continuous junction.
In this sense, our observation of the $m>1$ steps is
due to the discreteness of the system.

We also note that, in some parameter regime,
the resonating modes may be sub-harmonics of 
the fundamental frequency.
Sub-harmonic resonances of order $n$ can be sought by
generalizing (\ref{fourierexpansion}) to
\begin{equation}
  \psi(x) =  \sum_{m=-\infty}^{\infty} A_m e^{imx/n}  .
\label{subharmoicexpansion}
\end{equation}  
This is still a traveling wave but has an $n$ times longer period.
In a similar manner to the above discussion, one may expect resonances when
\begin{equation}
  0 = \omega^2_{m,n} - (m \omega/n)^2 \quad \mbox{i.e.} \quad
  \omega = n \omega_{m,n}/m
\label{subharmonicresonance}
\end{equation}
where
\begin{equation}
  \omega_{m,n} = 2 \Lambda_J \sin (2m\pi f/n) .
\label{subarmonicdispersion}
\end{equation}
The resonances with $m=1$ and $n=1,2,3$ have been 
observed by Caputo {\it et al.} \cite{lilli}.

\subsection{Amplitudes at resonances}

There was no need to analyze the coupling terms $F_m$
in the previous section to determine the resonant voltages, 
but it becomes necessary
when the AC components $A_m$ are wanted.
These mode amplitudes are important information 
for possible oscillator applications because one can then estimate the
power available from the device.

One way to calculate the amplitudes is to linearize the
coupled modal equations (\ref{modaleqns}) with respect to $A_m$.
One would obtain a linear algebraic system with coupling
between neighboring modes only.
We have tried this approach,
but the estimated magnitudes became too large to justify
the linearization,
and they did not agree well with the numerically obtained 
mode amplitudes.

Therefore, we have used another approximation
that can cope with larger amplitudes.
Since analysis of the modal equations in general
appears formidable,
we focus only on the steps $m=1$ and 2,
and make several approximations.

First, as we see in the spectra in Fig.~\ref{spectra}(a,b),
$A_3$ and higher modes have much smaller magnitudes
on these two steps, so they will be neglected.
Then, 
\begin{eqnarray}
  \phi(x) & = & x + (A_1 e^{ix} + \mbox{c.c.})
              + (A_2 e^{2ix} + \mbox{c.c.}) \nonumber \\
          & = & x + a_1 \sin (x+\theta_1) + a_2 \sin (2x + \theta_2)
\label{truncatedsolution}
\end{eqnarray}
where
\begin{equation}
  A_m = -(i/2) a_m e^{i \theta_m} .
\label{amplitudeconversion}
\end{equation}
The {\em harmonic balance} method, 
which is commonly used for approximating a finite
amplitude solution,
would neglect $A_2$ as well.
Thus, the ansatz (\ref{truncatedsolution}) can be thought of 
an extension of the method to a multi-harmonic case.
Including two harmonics, however, introduces coupling terms between them,
and makes the problem substantially harder.
By the Fourier-Bessel expansion, the coupling is expressed as
\begin{equation}
  F_0 = \sum_{n=-\infty}^\infty
    J_{-(2n+1)}(a_1) J_n (a_2) \sin \left[
    n \theta_2 - (2n+1)\theta_1 \right]
\label{F0}
\end{equation}
and
\begin{eqnarray}
  F_m & = & -(i/2) \sum_n
    J_{m-(2n+1)}(a_1) J_n(a_2) 
    e^{i(n \theta_2 + (m - 2n-1) \theta_1)}
  \nonumber \\
  & + & (i/2) \sum_n 
    J_{-m-(2n+1)}(a_1) J_n(a_2) 
    e^{-i(n \theta_2 - (m + 2n+1) \theta_1)}
\label{Fm}
\end{eqnarray}
for $m \neq 0$, and $J_m$ are Bessel functions.

To make analytical progress
we simplify the coupling
by taking only the most dominant terms (for small $a$'s)
and neglecting all the others.
This results in
\begin{equation}
  F_1 = -(i/2) J_0 (a_1) J_0 (a_2) , \quad
  F_2 = -(i/2) J_1 (a_1) J_0 (a_2) e^{i \theta_1} .
\label{F1F2}
\end{equation}
The higher order Bessel functions
$J_2(a_1)$, $J_3(a_1)$, $\dots$ and
$J_1(a_2)$, $J_2(a_2)$, $\dots$,
are neglected, which can be justified in the range, 
say, $|a_1| < 2$ and $|a_2| < 1$.
We substitute them into (\ref{modaleqns}) for $m=1,2$
and obtain
\begin{equation}
  (J_0 (a_1)/a_1)^2 = 
    (\delta_1^2 + \Gamma^2 \omega^2)/J_0^2(a_2) , \qquad
  \tan \theta_1 = -\Gamma \omega / \delta_1
\label{mode1}
\end{equation}
and
\begin{equation}
  (J_0 (a_2)/a_2)^2 = 
    (\delta_2^2 + 4 \Gamma^2 \omega^2)/J_1^2(a_1) , \qquad
  \tan (\theta_2-\theta_1) = -2 \Gamma \omega / \delta_2.
\label{mode2}
\end{equation}

It is convenient to think of $\omega$ as the control parameter
and determine $a_1$, $a_2$, $\theta_1$ and $\theta_2$.
The arguments of tangent are defined between 0 and $\pi$.
As $\omega$ increases from 0, 
$\delta_1$ and $\delta_2$ decrease monotonically,
and changes sign from positive to negative
at the resonance frequencies.
Thus, $\theta_1$ decreases from $\pi$ to 0,
with the crossover $\theta_1=\pi/2$ at $\omega=\omega_1$.
Similarly, $(\theta_2-\theta_1)$ decreases from $\pi$ to 0,
and crosses $\pi/2$ at $\omega=\omega_2/2$.

The amplitude equations of (\ref{mode1},\ref{mode2}) are more complicated.
It follows from (\ref{mode2}) that $a_2=0$ when $a_1=0$.
That is, the second mode is excited only in the presence of the first mode.
On the other hand,
the first mode can be excited alone since $a_1>0$ in (\ref{mode1})
when $a_2=0$.
The feedback from the second mode through $J_0(a_2)$ 
in (\ref{mode1}) is not essential
when considering $a_1$.

To extract more information we make a further approximation.
We first linearize the Bessel functions on the right hand sides,
i.e.\ $J_1(a_1) \approx a_1/2$ and $J_0(a_2) \approx 1$.
(Consequently, $F_1$ becomes independent of $A_2$,
and the feedback from the second mode is neglected.
Then, (\ref{mode1}) reduces to the equations obtained through
the method of harmonic balance \cite{jap96}.)

On the left hand sides of (\ref{mode1},\ref{mode2}),
the function $J_0(a)/a$ is monotonically
decreasing from $\infty$ to zero as $a$ increases from 0 to
the first zero of $J_0$, which is $a^* \approx 2.4$.
Thus, for any $\omega$, $a$ can be found within 0 and $a^*$.
Approximating the function crudely by $J_0(a)/a \approx 1/a$
would make the range of $a$ unbounded.
This is not desirable  near the resonances.
We could include the next term from the expansion of $J_0$,
but this would make the following algebra more complicated.
Alternatively, we replace the function by another one
which has the same asymptotic behavior as $a \rightarrow 0$
and a bounded domain for $a$.
We use
\begin{equation}
  (J_0(a)/a)^2 \approx 1/a^2 - 1/4 .
\label{lhsapprox}
\end{equation}
This has a slightly different saturation amplitude at $a^*=2$
but overall the approximation is good;
the error in $a$ introduced by this replacement is
less than 5\% for $a<0.7$
and at most 20\% when $a$ saturates.
In return for the error introduced at this stage,
we obtain simple expressions for the amplitudes as
\begin{equation}
  a_{1} = ( b_{1}^2 + 1/4 )^{-1/2} \qquad {\rm and} \qquad 
  a_{2} = ( b_{2}^2 + 1/4 )^{-1/2} ,
\label{amplitudeestimate}
\end{equation}
where
\begin{equation}
  b_1^2 = \delta_1^2 + \Gamma^2 \omega^2 \qquad {\rm and} \qquad
  b_2^2 = 4 (\delta_2^2 + 4 \Gamma^2 \omega^2)/a_1^2 .
\label{amplitudeestimatesupplement}
\end{equation}

Once the phases and amplitudes of the modes are determined, 
the instantaneous voltage is obtained from 
the time-derivative of (\ref{truncatedsolution}).
In our normalization,
\begin{equation}
  V / I_cR_n = \Gamma \omega \left[ 1 +
  a_1 \cos(x+\theta_1) + 2 a_2 \cos(2x+\theta_2) \right].
\label{truncatedvoltage}
\end{equation}
Thus, the DC voltage and the AC amplitude of the mode $m$ are
\begin{equation}
  V_{\mbox{\scriptsize DC}}/I_cR_n = \Gamma \omega \qquad {\rm and} \qquad
  V_{\mbox{\scriptsize AC}}/I_cR_n = \Gamma \omega m a_m .
\label{DCandACvoltages}
\end{equation}

\begin{figure}[tbp]
\centerline{
\psfig{file=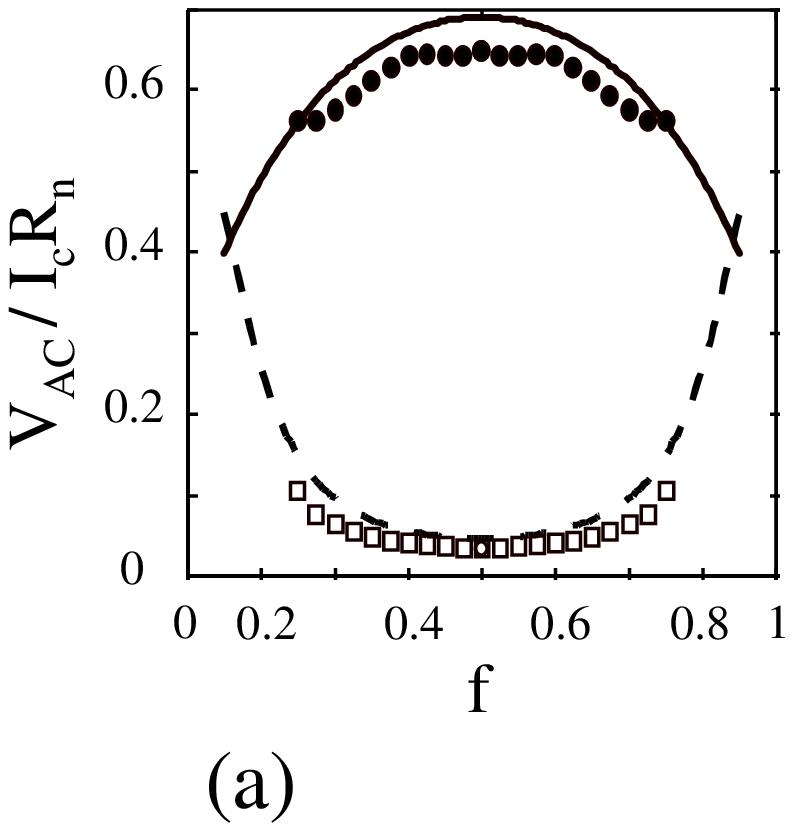,height=2.5in,width=2.5in}
\psfig{file=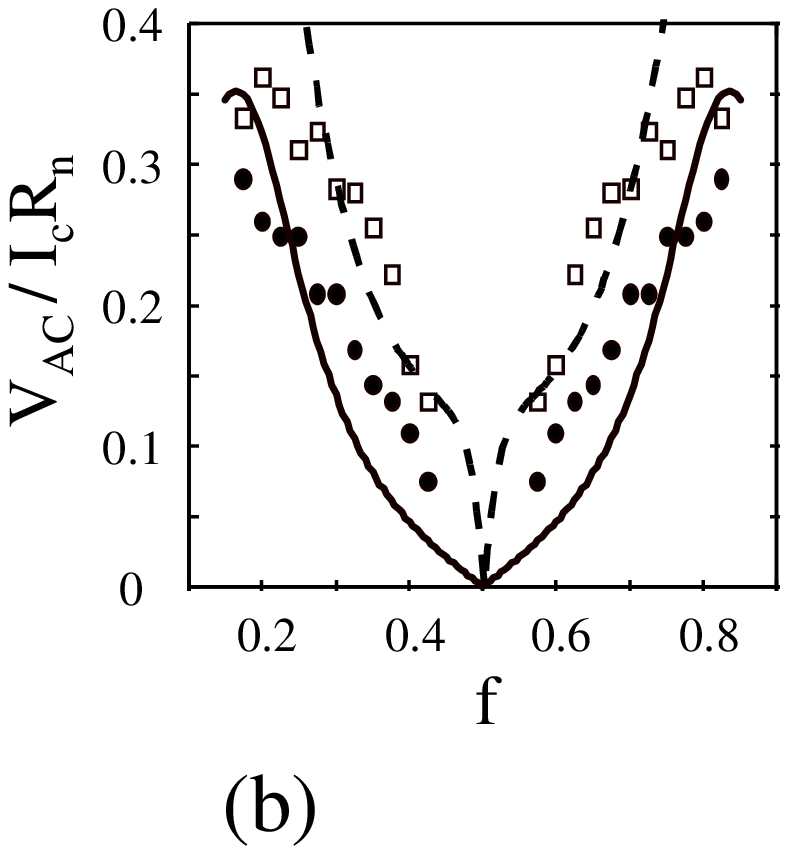,height=2.5in,width=2.5in}
}
\caption{Amplitudes of the first two AC voltage harmonics 
on (a) the $m=1$ (Eck) step and (b) the $m=2$ step vs.\ $f$.
The theory (25) is represented by a solid (dashed) line 
for the first (second) harmonic.
Filled circles and open squares show the amplitude
of the first and second harmonics, respectively, from simulations.
The values of $\Lambda_J$ and $\Gamma$ are the same as 
in Fig.~\protect{\ref{fig:simIV}}.
As expected, the amplitude at the second frequency is quite small
on the $m=1$ step (a), while the amplitudes of the first two harmonics
are comparable on the $m=2$ step (b). 
}
\label{amplitudes}
\end{figure}

In Fig.~\ref{amplitudes}, we compare the amplitude estimate
with simulations.
On the first two steps, 
the peak AC voltage amplitudes for the first two Fourier modes 
are obtained from simulations, shown as data points.
The estimates of the resonance frequencies
$\omega=\omega_1$ and $\omega_2/2$ 
from (\ref{resonancefrequency1}) are used.
The theoretical curves show a reasonable agreement with the data
on the Eck step.
On the second step the comparison is worse,
but the overall magnitude and $f$-dependence are estimated fairly,
considering the several approximations made during the derivation.

The maximum AC voltage oscillation of the first mode 
on the Eck step is achieved at $f=0.5$.
The maximum of the second mode on the second step
happens near the lower critical frustration ($f \approx 0.2$).
The vertical axes in Fig.~\ref{amplitudes} are shown
in terms of AC voltages.
Converted back to the amplitudes, 
$a_1<1.75$ and $a_2<0.4$ on the Eck step for $0.2<f<0.8$.
This is thought to be within the validity of the assumptions.
On the second step, 
$a_1<1.8$ for $0.15<f<0.45$,
but $a_2$ exceeds unity when $f<0.25$.
The bending of the theoretical curves in (b) when $f<0.25$
may be attributed to the large predicted magnitude of $a_2$.
Including terms involving $J_1(a_2)$ in the modal equations
would be necessary to make a better estimate in the region.

Generally, our assumption is violated when $a_1$ saturates
in (\ref{amplitudeestimate}), or when $b_1^2$ becomes small.
On the Eck peak $b_1^2 \approx (2 \Gamma \Lambda_J \sin (\pi f))^2$,
so the assumption breaks down for very small $\Gamma$ and/or $\Lambda_J$.
Similarly, $a_2$ exceeding unity indicates that we need more
coupling terms included from (\ref{modaleqns}).
On the $m=2$ step, $a_2<1$ is fulfilled when
$\Gamma^2 \omega_2^2 [1+\Gamma^2 \omega_2^2+
4 (\omega_1^2-\omega_2^2/4)^2]>3/4$.
Again, small $\Lambda_J$ and/or $\Gamma$  make it
easier to violate the criterion.
Thus, the validity of our approximations is limited 
to when $\Gamma \Lambda_J$ is relatively large
and resonances are not so strong.

\subsection{Power at resonances}

The calculation of the mode amplitudes provides a theoretical
upper limit 
for the power available from the Josephson oscillator at resonance.
If we assume a matched load condition, then the power
dissipated in the load is just the power dissipted in the matched
resistance.  
Using the RSJ model and assuming sinusoidal voltage oscillations,
this correponds to 
$P= V_{AC}^2 / 8 R_{arr}$, where $R_{arr}=R_n/N$ is the
array resistance.  
The power available from mode $m$ is then 
\begin{equation}
  \frac{P_m}{I_c^2 R_n} = \frac{N}{8} \left(\frac{V_{AC}}{I_c R_n}\right)^2
  = \frac{N}{8} \left(\Gamma \omega m a_m\right)^2
\label{powerestimate}
\end{equation}
with $a_m$ estimated in (\ref{amplitudeestimate}).
Therefore, the vertical axis of Fig.~\ref{amplitudes} 
is also $\sqrt{8 P_m/N I_c^2 R_n}$, 
and the $f$-dependence of the powers can be read directly from the figure.
When biasing on the Eck step, the largest power
and the highest frequency are obtained at $f=0.5$.  
When the system is biased on the second step, 
the largest power is obtained near $f=0.2$,
and the highest frequency is achieved near $f=0.25$.

We take our experimental values ($I_cR_n=0.93 \,$mV and $R_n=16.6 \, \Omega$)
and the maximum amplitudes in Fig.\ref{amplitudes}
($V_{AC,1} = 0.65 I_c R_n$ on the Eck step and 
$V_{AC,2} = 0.36 I_c R_n$ on the second step).
This predicts a maximum of about $150$~nW from the first harmonic 
at the first resonance and $46$~nW from the second harmonic 
at the second resonance.
These numbers are much smaller than previous estimates \cite{benz}, and,
according to measurements \cite{Ust:LJJ,booi}, are probably more realistic.

Our simulation and analysis show that 
multiple Fourier modes are excited on the new steps $m>1$.
For possible applications as oscillators,
this is not desirable since the AC power is therefore distributed among
the modes, instead of being concentrated in one mode, as at the Eck step.
Usually in such a case one could simply increase the drive, to increase
the output at the desired frequency.  However, in this system the driving
current is limited because if it is increased beyond the top of the step,
the resonance becomes unstable. Furthermore, since the maximum output power
is small for a single row, methods are needed to combine power
from multiple rows while preserving the frequency content of the resonance.

Both discrete rows of underdamped Josephson junctions and continuous
long junctions operating at the Eck step have already been proposed as
high frequency oscillators \cite{Herre:Eck,Ped:rev}. 
Although no single experiment has directly
compared the two systems, studies indicate that the ouput power levels are
comparable, while the output impedence of the discrete system may be 
advantageously higher \cite{Herre:Eck}.    In order to increase power levels in
either case, additional
oscillators are required.  Stacks of continuous long junctions have been
fabricated and measured \cite{Cara,Ust:st,Sakai}, 
as well as the analogous inductively 
coupled discrete arrays \cite{jap96,asc96}.
However, in both systems, the power is only increased if the  oscillating
elements are phase-locked and in-phase.  
The existence of in-phase as well as anti-phase states have been 
well established for both discrete and continuous
systems, through numerics \cite{GJ}, simulations \cite{Petr,SBP}, 
and experiments \cite{jap96,monaco}.
Unfortunately, in both cases the
anti-phase pattern is preferred at the Eck step \cite{jap96,monaco},  
and the output oscillations of adjacent
elements almost completely cancel.  In addition, the difficulty (and 
necessity) of 
fabricating identical oscillators in the case of continuous stacks limits
the possibility for improving this system.

On the other hand, the presence of higher harmonic resonances in discrete
systems may actually
be an advantage in terms of power combining.  As observed in \cite{asc96}, 
when the discrete system is biased on the stable $m=2$ peak,
the fundamental oscillations of the rows are anti-phase (shifted by $\pi$),
but the second harmonics are still in-phase.
Thus, the output power from the second mode is enhanced.
It is important to note that such modes do not exist in uniform continuous
long junctions.
So although the power level of a single row biased in this state is lower
than the oscillators mentioned above, the potential for power-combining
in this state is much greater. 
Simulations indicate that this in-phase resonance 
for the second mode is stable for many coupled rows \cite{asc96}, 
which is promising for oscillator applications.

\section{Summary}

We have observed experimentally and through simulations 
new resonance steps in long one-dimensional Josephson junction arrays
in the underdamped and discrete regime.
They emerge in the $I$--$V$ characteristic 
below the Eck voltage,
and are tunable in the magnetic field $f$.
We have shown that the resonance mechanism can be explained by
neglecting the boundary effects
and assuming a traveling wave solution.
Fourier components of the solution resonate to create
a waveform which appears to consist of a kink and radiation,
phase-locked to each other.
We have employed a two-mode extension of the harmonic balance method
in order to simplify the coupling of the modes.
We have derived an analytic formula for not only 
the resonance voltages but also the amplitudes 
of the modes. Finally, we used these results to  
predict the power available
from such Josephson oscillators and discuss its dependence on the system
parameters.

\section*{Acknowledgment}
We thank Mauricio Barahona, Alexey Ustinov, Andreas Wallraff, 
and Pasquilina Caputo for valuable discussions. 
We acknowledge the support of the NSF Graduate Fellowship program and
NSF Grants DMR-9402020, DMS-9057433, and DMS-9500948.  We appreciate the
partial support of the Dutch Foundation for Fundamental Research on
Matter.

\end{document}